\newif\ifpreprint

\preprintfalse
 
\ifpreprint

	\documentclass[aps,pra,superscriptaddress,preprint]{revtex4-1}
\else

	\documentclass[aps,pra,superscriptaddress,reprint,twocolumn,showpacs,longbibliography ]{revtex4-1}
\fi

\usepackage{amsmath,epsfig,color,amssymb}
\usepackage{graphicx}
\usepackage{dcolumn}
\usepackage{bm}
\usepackage{amsmath, amsfonts, amssymb,mathrsfs}
\usepackage{pstricks}
\usepackage{amsxtra}
\usepackage{amsthm}
\usepackage{natbib}
\usepackage{siunitx}
\usepackage{float}

\usepackage[normalem]{ulem}
 \usepackage{soul}

\definecolor{ForestGreen}{rgb}{0,0.5,0}

\begin{document}

\title{Observation of chiral photocurrent transport in the quantum Hall regime in graphene    }

\author{Olivier Gazzano}
\thanks{These two authors contributed equally}
\affiliation{Joint Quantum Institute, NIST/University of Maryland, College Park, Maryland 20742}
\author{Bin Cao}
\thanks{These two authors contributed equally}
\affiliation{Joint Quantum Institute, NIST/University of Maryland, College Park, Maryland 20742}
\affiliation{Department of Physics, University of Maryland, College Park, Maryland 20742}
\author{Jiuning Hu}
\affiliation{NIST, Gaithersburg, Maryland 20878}
\author{Tobias Huber}
\affiliation{Joint Quantum Institute, NIST/University of Maryland, College Park, Maryland 20742}
\author{Tobias~Grass}
\affiliation{Joint Quantum Institute, NIST/University of Maryland, College Park, Maryland 20742}
\author{Michael~Gullans}
\affiliation{Joint Quantum Institute, NIST/University of Maryland, College Park, Maryland 20742}
\affiliation{Department of Physics, Princeton University, Princeton, New Jersey, 08544}
\author{David Newell}
\affiliation{NIST, Gaithersburg, Maryland 20878}
\author{Mohammad Hafezi}
\email{hafezi@umd.edu}
\affiliation{Joint Quantum Institute, NIST/University of Maryland, College Park, Maryland 20742}
\affiliation{Department of Physics, University of Maryland, College Park, Maryland 20742}
\affiliation{Department of Electrical and Computer Engineering and IREAP, 
 University of Maryland, College Park, Maryland 20742}
\author{Glenn S. Solomon}
\email{gsolomon@umd.edu}
\affiliation{Joint Quantum Institute, NIST/University of Maryland, College Park, Maryland 20742}
\affiliation{Department of Physics, University of Maryland, College Park, Maryland 20742}
\affiliation{NIST, Gaithersburg, Maryland 20878}


\begin{abstract}
Optical excitation provides a powerful tool to investigate non-equilibrium physics in quantum Hall systems. Moreover, the length scale associated with photo-excited charge carries lies between that of local probes and global transport measurements. Here, we investigate non-equilibrium physics of optically-excited charge carriers in graphene through photocurrent measurements in the integer quantum Hall regime. We observe that the photocurrent oscillates as a function of Fermi level, revealing the Landau-level quantization, and that the photocurrent oscillations are different for Fermi levels near and distant from the Dirac point. Our observation qualitatively agrees with a model that assumes the photocurrent is dominated by chiral edge transport of non-equilibrium carriers. Our experimental results are consistent with electron and hole chiralities being the same when the Fermi level is distant from the Dirac point, and opposite when near the Dirac point.

\end{abstract}
\date{\today}

\maketitle

The integer quantum Hall effect observed in graphene reflects its gapless relativistic band structure at low energy~\cite{Zhang2005b,Novoselov2005,Geim2007}; for instance, the observation of a Landau level at the Dirac point and an anharmonic Landau level spacing. This permits optical transitions only between specific Landau levels~\cite{Orlita2011,Jiang2007,Chen2014,Sadowski2006}. Recently, optical probing has been used to study defects~\cite{Nazin2010b}, photovoltaic effect~\cite{Masubuchi2013a}, thermal properties~\cite{Cao2016,Wu2016}, and carrier multiplication and relaxation~\cite{Sonntag2017,Wendler2014,Plochocka2009}. The photocurrents generated through optical excitation are due to non-equilibrium (hot) carriers transport of both carrier types (electrons and holes)~\cite{Nazin2010b}. Such optical measurements allows one to perform a relatively local measurement of quantum Hall states, on the scale of the wavelength of light. This is complementary to high-resolution measurements, such as scanning-tunneling spectroscopy which probes the local density of states~\cite{Loftus2007,Brar2007,Hashimoto2008,Ghahari2017}, and global transport measurements.

As in other quantum Hall systems,  the confining  potential on the boundaries of  the  graphene sample leads  to  chiral  transport  of  carriers  through edge  states~\cite{Halperin1982,DasSarma2011}. For electrons  and  holes  occupying  states on the same side of the Dirac point, the  edge-state confinement predicts  that  they  share  the  same  chirality~\cite{Abanin2007,Queisser2013}. This prediction has been used to describe supercurrent transport measurements in a graphene-superconductor  interface~\cite{Hoppe2000,Amet2016,Lee2017}. However, electrons  above the Dirac point and holes below the Dirac point have opposite edge-states chiralities. It is intriguing to investigate the manifestation of this chirality switching using optical excitation of electrons and holes in the non-equilibrium regime by adjusting the Fermi level. 

In this paper, we demonstrate such chirality by measuring the photocurrent in the integer quantum Hall regime. We use near-infrared light to excite electrons from below the Dirac point to states above the Dirac point, Fig.~1(a),(b). We measure the resulting photocurrent and distinguish between bulk and edge contributions of both electrons and holes. We correlate oscillations in edge-state photocurrent with changes of the density of states at the Fermi level. With the Fermi level at the Dirac point, electrons and holes in edge states propagate in opposite directions resulting in maximum edge-state photocurrent. With the Fermi level well above or well below the Dirac point, we show that electrons and holes in edge states propagate in the same direction giving rise to two photocurrent polarity changes when sweeping the Fermi level across one Landau level.

\begin{figure*}
\begin{center}
\includegraphics[width = 0.9\linewidth]{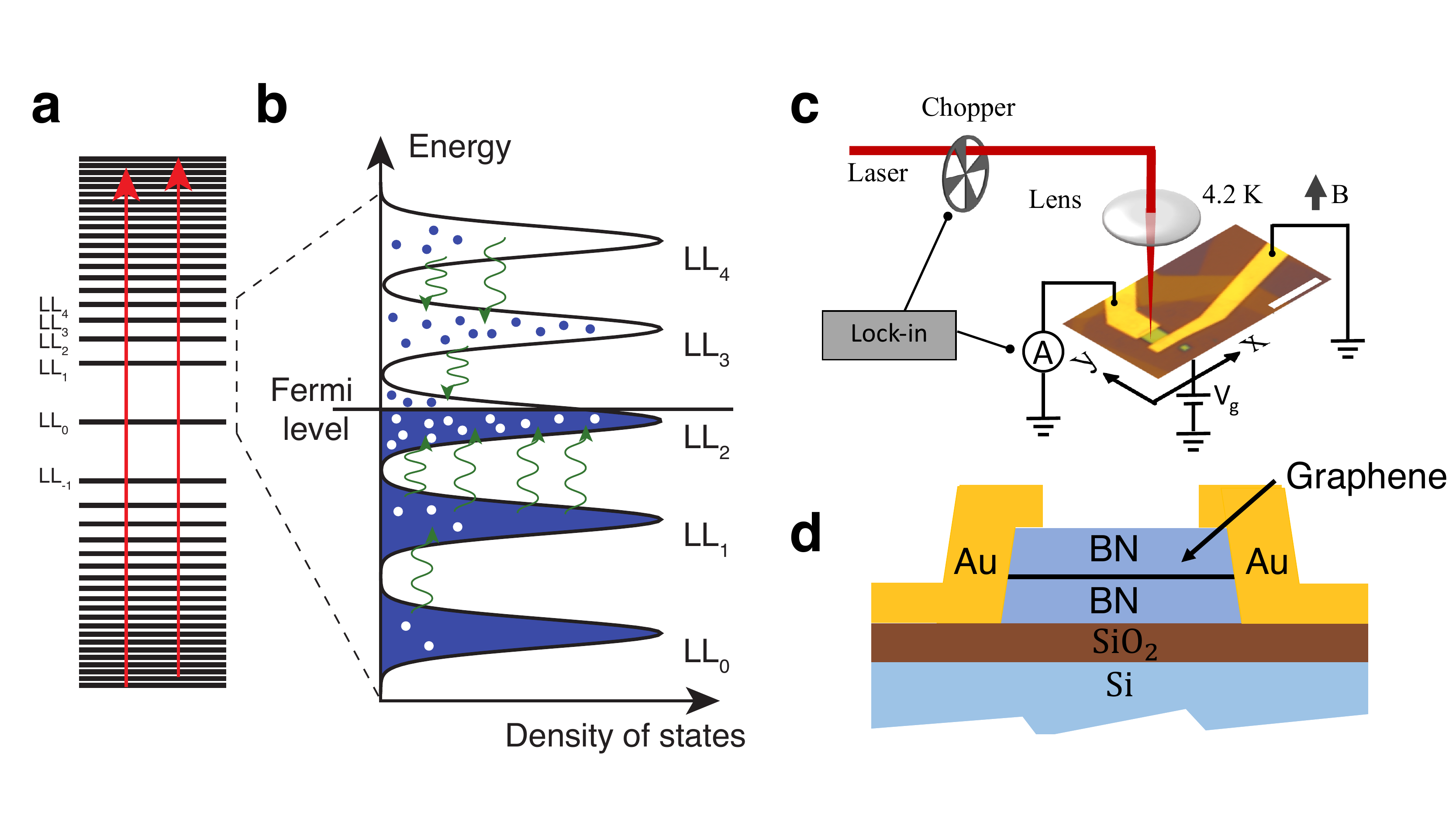}
\caption{\textbf{Landau level quantization of graphene density of states and experimental setup}. \textbf{(a)}. Schematic showing Landau level  quantization. Red arrows represent optical pumping of allowed transitions $LL_{n} \rightarrow LL_{n'}$, such that $n+n'=\pm1$ where $n, n'$ are the principle quantum numbers of the Landau levels. \textbf{(b)}. Photo-excited hot electrons (solid blue circles) and holes (open circles) relax to the Fermi level $E_F$. Blue indicates occupied electron sea. $E_F$ is above the half-filling of $LL_2$, resulting in unbalanced relaxation of hot-carriers: holes outnumber electrons in $LL_2$, leading to an electron dominated hot carrier population.  \textbf{(c)}. Schematic of the experimental setup and optical image of the graphene sample with metallic gold contacts (white scale bar: 10~$\mu$m). We define the \textit{x}-direction as along the metallic contact edge. A lock-in measurement is triggered by the chopping frequency of the laser. The sample temperature is 4.2 K. \textbf{(d)}. Schematic showing cross section of the boron-nitride (BN) encapsulated graphene sample shown in \textbf{(c)}. The doped Si acts as the back gate. For more details, refer to the Appendix.}

\label{fig:fig1nn}
\end{center}
\end{figure*}

\ifpreprint \newpage  \fi

\begin{figure*}
\begin{center}
\includegraphics[width = 1\linewidth]{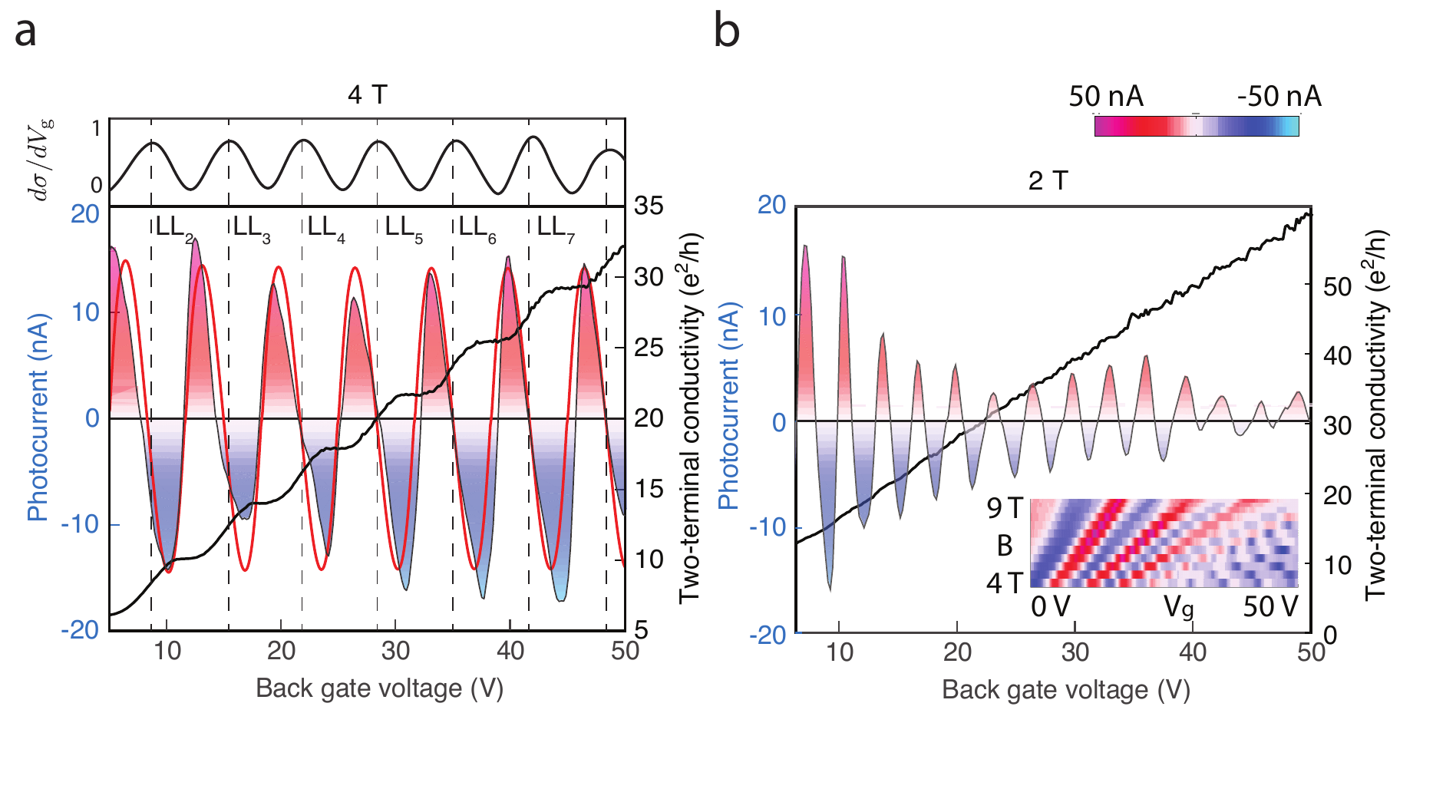}
\caption{\textbf{Oscillations of the photocurrent induced by the Landau quantization in the device.} \textbf{(a).} Right: Two-terminal conductivity (black line) in the quantum Hall regime (4~T) as a function of the back gate voltage. The Dirac point corresponds to a back gate voltage of $-7~$V. Left: Photocurrent generated by the laser on the sample edge as a function of the back gate voltage. Strong oscillations of the photocurrent are measured and zeros of photocurrent correspond to half-fillings (dashed lines) and integer-fillings of Landau levels for $LL_{n\geq4}$. Photocurrent data is in good agreement with our calculation (red line). \textbf{(b).} Two-terminal conductivity (black lines, right scale) and photocurrent (left scale) as a function of the back gate voltage at 2~T. The quantum Hall plateaus in the two-terminal conductivity are not visible while the photocurrent oscillations remain distinct. The inset shows the photocurrent as a function of back gate voltage measured at various magnetic fields. }
\label{fig:fig2nn}
\end{center}
\end{figure*}

A schematic of our experimental setup and device structure are shown in Fig.~\ref{fig:fig1nn}(c),(d). An optical microscope image of the graphene sample (2.49~$\mu$m by 3.87~$\mu$m) is shown with metallic gold contacts on two sides. These two contacts are unbiased in photocurrent measurements. 
The exfoliated graphene layer is sandwiched between exfoliated boron nitride, and the structure is back-gated using the Si substrate (See Appendix for details). The carrier density (\textit{i.e.} Fermi energy) in the sample can be tuned by changing the back gate voltage~\cite{A.A.2004, Wang2008}. The photocurrent data are taken at 4.2~K predominantly with an out-of-plane magnetic field of 4~T. Additional data at a magnetic field of 9~T magnetic field is shown in the Appendix. Except where noted, the light source is a laser tuned to 930~nm with a power fixed at 10~$\mu$W. The laser spot size on the sample is 1.8~$\mu$m. 

Fig.~\ref{fig:fig1nn}(a),(b) schematically show the Landau level quantization of the graphene density of states and the laser excitation. In the integer quantum Hall regime, the optically allowed interband transitions are between Landau levels $LL_{n} \rightarrow LL_{n'}$ such that, $n+n'=\pm1$ where $n, n'$ are the respective principle quantum numbers, and are negative below and positive above the Dirac point~\cite{Sadowski2006,Jiang2007}. However, in the near infrared energy range, the transition energies between Landau levels are not expected to be discrete because of Landau level broadening at these high energies, in particular due to the sample impurity and disorder effects~\cite{Funk2015}. We measure the photocurrent between unbiased ohmic contacts (Fig.~\ref{fig:fig2nn}(a))~\cite{Nazin2010b,Masubuchi2013a,Cao2016,Sonntag2017}. By sweeping the laser wavelength in the ranges 0.9 to 1~$\mu$m ($|n|\approx36$) and 1.9 to 2.0~$\mu$m ($|n|\approx8$) at 9 T we find that the photocurrent is unchanged for all the wavelengths in each range, confirming the absence of well-defined Landau levels in this energy range. For a laser power of 10~$\mu$W and at a magnetic field above 2~T, the maximum of photocurrent we measured is 30~nA. This value corresponds to approximately 10~\% of the carriers generated by the laser (see Appendix). Only small photocurrents ($<2$~nA) are measured at zero magnetic field (see Appendix). 

To demonstrate that the photocurrent reveals the integer quantum Hall physics, we measure the photocurrent generated by the laser while sweeping the back gate voltage. Fig.~\ref{fig:fig2nn}(a) shows that the photocurrent oscillates as a function of the back gate voltage under a magnetic field of 4~T when the laser is 1.5~$\mu$m away from the sample center. To show that the photocurrent oscillations are correlated with the Landau levels, we perform a two-terminal conductance measurement, as shown in Fig.~\ref{fig:fig2nn}a (black line). The conductance measurement exhibits integer quantum Hall plateaus \cite{Novoselov2005,Zhang2005b} correlated with photocurrent oscillations. In particular, away from the Dirac point (for Landau levels $LL_{n\geq4}$), the zeros of the photocurrent correspond to both the middle of each plateau and the middle of the transition between two plateaus. These points corresponds to full filling and half filling of Landau levels, respectively. An alternative way to connect filling factors and changes in photocurrent polarity is to plot the derivative of conductance measurement with respect to the back gate voltage~$V_{\textrm{g}}$, as shown in the upper panel of Fig.~\ref{fig:fig2nn}(a). In inserts in Fig.~\ref{fig:fig2nn}(b) and the Appendix Fig.~\ref{fig:S2nn},  we also change the magnetic field and observe the same correlation between photocurrent and conductance measurements.

The photocurrent measurements are more sensitive in the low magnetic field regime than the two-terminal transport measurements. Fig.~\ref{fig:fig2nn}(b) shows  the two-terminal conductivity and the photocurrent as a function of the back gate voltage, at a low magnetic field of  2~T. While the quantum Hall plateaus are not visible in the two-terminal measurement, the oscillations of the photocurrent are pronounced. One explanation is that the two-terminal transport measurement evaluates the sum of edge state conductance, while the photocurrent is the difference of two components currents (electrons and holes) making it more sensitive.

The photocurrent oscillations track the back gate voltage, indicating that the physics is influenced by the density of states near the Fermi level. However, the polarity of the current indicates that the contributing carriers are not at the Fermi level, but are non-equilibrium (hot) carriers. To further clarify this, we set the Fermi level slightly above half-filling of a Landau level, as illustrated in Fig.~\ref{fig:fig1nn}(b). In this regime,  the number of available hole states in the Landau level is larger than that of the electron states. If we assume relaxation to available Fermi-level states is fast~\cite{Plochocka2009}, then the number of holes  that relaxed to the Fermi level is larger than that of the electrons. The transport of the carriers near the Fermi level would be hole dominated, while hot carrier transport would be electron dominated. In this case, the measured polarity of the photocurrent indicates that the transport to the contact is dominated by electrons. Thus, we conclude that the photocurrent is due to hot carriers, and not the carriers in the vicinity of the Fermi level, as would be the case for local heating.

We develop a model to explain the observed dependence of the photocurrent on the back gate voltage. In the model, photocurrents are due to hot carriers, as described above. We further assume that the photocurrent is dominated by the edge physics, and carriers reach the edges with a probability set by the laser spot location relative to the sample edges. The validity of this is discussed later.

\begin{figure*}
\begin{center}
\includegraphics[width = 1\linewidth]{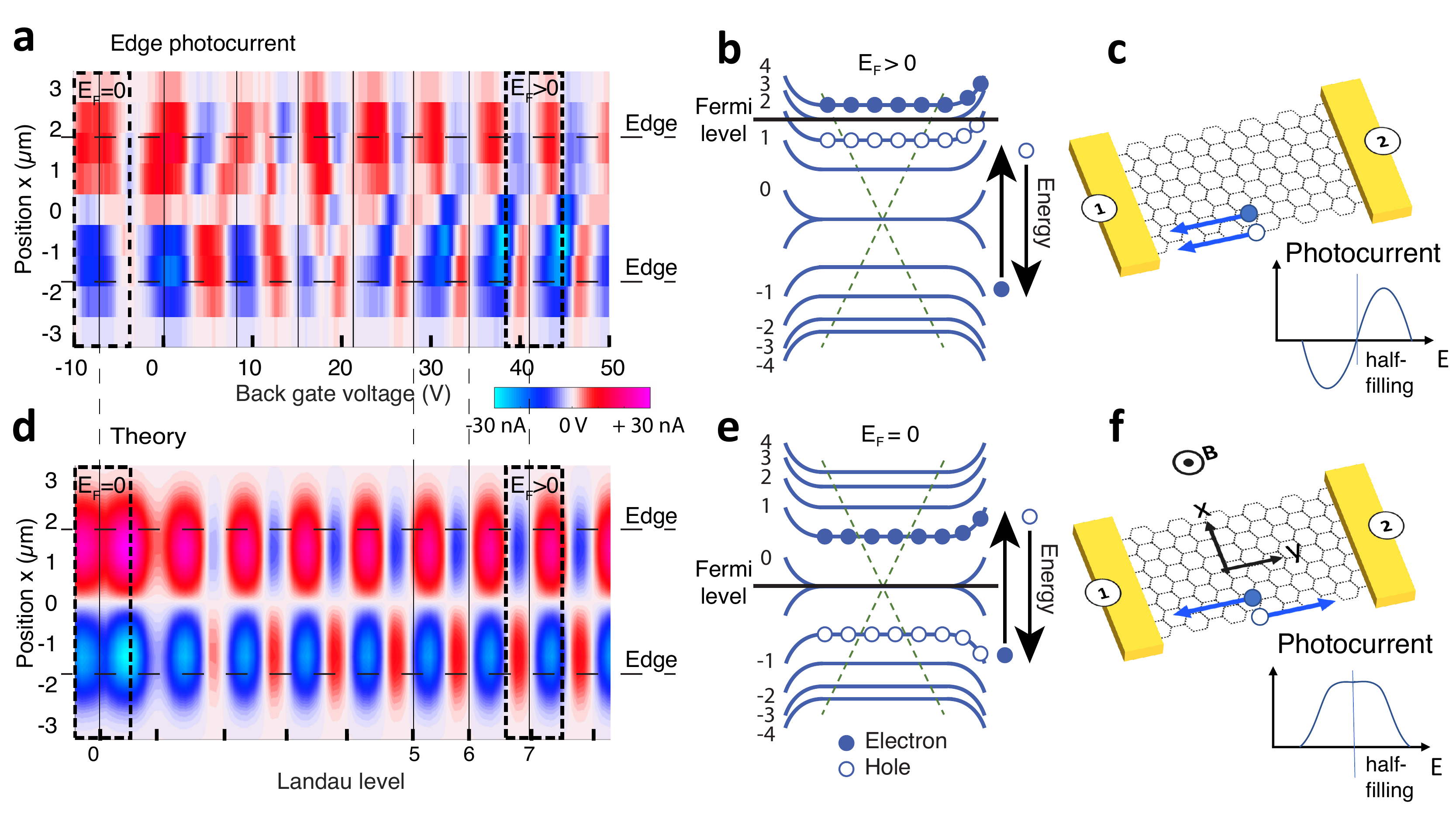}
\caption{\textbf{Chiral transport of the photo-carriers.} \textbf{(a).} Edge photocurrent as functions of back gate voltage for different laser positions ($I_{\text{edge}}=I_{+ \textbf{B}}-I_{- \textbf{B}}$, $B=4$~T). We move the laser spot along the x-direction on the sample, as shown in Fig.~\ref{fig:fig1nn}~(c). Photocurrent polarity oscillates between red and blue with respect to $V_{\textrm{g}}$ while the polarity swaps when the laser position scans from one edge to the other edge of the graphene sample. The horizontal dashed lines represent the positions of sample edges. The half-fillings (vertical black lines, extracted from the two-terminal transport measurement) match the zeros of the photocurrent when the Fermi level is away from the Dirac point ($V_{\textrm{g}} > 20~V$ or $LL_{n\geq4}$), while at the zeroth Landau level, the half-filling (the Dirac point) coincides with the center of a photocurrent peak. When the Fermi level is away from the Dirac point ($V_{\textrm{g}} > 20~V$ or $LL_{n\geq4}$) as shown in \textbf{(b)}, hot electrons and hot holes are on the same side of the Dirac point and the chiralities for electrons and holes are the same, as in \textbf{(c)}, leading to a polarity change at the half-filling of the Landau level. When the Fermi energy is in the vicinity of the Dirac point $E_F\simeq 0$, hot electrons and holes are separated on different sides of the Dirac point, as is shown in \textbf{(e)}, and the chiralities for electrons and holes are different, as shown in \textbf{(f)}, making the photocurrent not changing polarity at the zeroth Landau level. \textbf{(d).} Simulation of the edge state photocurrent. }
\label{fig:fig3nn}
\end{center}
\end{figure*}

To determine the direction of the edge current due to electrons and holes, we now discuss their behavior in the presence of the confining edge potential. By solving the Schr\"odinger equation in the Landau gauge $\mathbf{A}=xB\mathbf{y}$, the system is translationally invariant along the $y$-axis, and the energy spectrum as a function of the confinement potential $V(x)$ is $
\epsilon_{\lambda,n}=\lambda\frac{\hbar v_F}{l_B}\sqrt{2n}+\lambda \, \textrm{sgn}(q) V(x),$ where $\lambda$ is the band index, $+1$ for conduction band and $-1$ for valence band. $v_F$ is the Fermi velocity, $l_B$ is the magnetic length, $n$ is the Landau level index, $q$ is the  carrier charge. Therefore, the group velocity in the \textit{y}-direction is given by:

\begin{equation}
v_{g}=\frac{\partial \epsilon_{\lambda,n}}{\partial k_y}\propto \lambda \, q\, \frac{\partial{V(x)}}{\partial{x}}
\label{eqn:eqn1}
\end{equation}
where $k_y$ is the edge state momentum in the \textit{y}-direction. This equation determines the direction of electron and hole edge transport. For example, above the Dirac point, since the confining potential and the charge have opposite signs for electrons and holes, their group velocities are in the same direction, and their currents are in opposite direction, as shown in Fig.~\ref{fig:fig3nn}(b),(c).

Using this picture, we explain the oscillation of the photocurrent with the back gate voltage. At both half- and integer-filling of Landau levels, the number of available electron and hole states are equal in the vicinity of the Fermi level. This makes the number of hot electrons and holes equal, leading to a net zero edge current. By moving away from half- and integer-filling factors, the number of available electron and holes are no longer equal; thus, the photocurrent becomes nonzero, and makes a full oscillation per Landau Level. 

In our numerical model, we sum over thermally occupied edge channels, for both electrons and holes, and we evaluate the resulting photocurrent. The details of the model are described in the Appendix. The result of this model is presented as a red curve in Fig.~\ref{fig:fig2nn}(a), which qualitatively agrees with our observation.  Specifically, the signal oscillates with the back gate voltage, changing polarity twice per Landau level, at half- and integer-fillings. Note that the discrepancy between photocurrent amplitudes of the two polarities on the same edge can be explained by the bulk mobility difference of electrons and holes, which is included in our numerical model.  

To study the spatial dependence of the photocurrent, we scan the laser spot position parallel to the contacts, \textit{i.e.} along \textit{x}-direction (see Fig.~\ref{fig:fig1nn}(c)). The photocurrent consists of two current components: one is diffusive, directly to the ohmic contacts while the other reaches the contacts through the edge states~\cite{Song}. The purely diffusive component is symmetric with the \textit{B}-field. The second component is chiral, and thus antisymmetric with the \textit{B}-field. Thus, the two components can be separated by taking the sum and difference of photocurrents, for +\textit{B} and -\textit{B} fields~\cite{Cao2016}. Fig.~\ref{fig:fig3nn}(a) shows the isolated magnetic-field dependent photocurrent as a function of $V_{\textrm{g}}$ and laser spot position (the original $\pm$ \textit{B} data is in Fig.~\ref{fig:S3nn} in the Appendix). We observe that the photocurrent difference is weak in the middle of the sample (position 0~$\mu$m), compared to the edges.  Moreover, the photocurrent polarities have opposite sign on the two edges of the sample. These two observations confirm that the difference of photocurrents, for +~\textit{B} and -~\textit{B} fields is due to the diffusion of carriers to the nearest edge state, after which they are transported to the contacts via the edge states. We note, in contrast to transport measurements where the edge current balance is broken by the application of an electric field, here, the current imbalance is due to fact that the laser spot is off-centered on the sample.

We observe that regardless of the excitation position of the laser in \textit{x}-direction, when the Fermi level is above the Dirac point ($E_F>0$), the photocurrent changes polarity twice per Landau Level, as explained above. However, when the Fermi level is in the vicinity of the Dirac point $E_F\simeq 0$, the photocurrent does not change polarity, as we sweep through the zeroth Landau level. This observation is also described by our model. Specifically, when the Fermi level is in the vicinity of the zeroth Landau level, electrons are above the Dirac point while holes are below the Dirac point leading to the same confining potential sign for electrons and holes. According to Eqn.~\ref{eqn:eqn1}, the group velocity is opposite for electrons and holes in this case, and their respective currents are therefore in the same direction, Fig.~\ref{fig:fig3nn}(e),(f). Thus, by sweeping through the zeroth Landau level, the photocurrent shows a maximum and does not change sign. A simulation based on our model is shown in Fig.~\ref{fig:fig3nn}(d) which qualitatively agrees with the experimental result. The simulation shows a small double peak at the zeroth Landau level due to fast electron-hole recombination, however this double peak is not resolved in the experiment.

We note our observation can not be explained by local heating (the photo-Nernst effect) which was previously discussed in Ref. ~\cite{Cao2016} and our theoretical model differs from it.   Broadly summarizing, the Nernst effect yields a current that is related to two adjacent regions at different equilibrium temperatures. If the graphene sample is heated, the transport is modulated as the Fermi-level occupation is locally altered~\cite{Zhang2005b, Zuev2009}. In contrast, in our model, the electron and hole populations are in nonequilibrium distributions. The transport is due to hot carriers created by the laser, with a spot size on the order of the sample size, and we see no evidence of laser heating when comparing two-terminal transport measurements with and without the laser. Furthermore, details of our photocurrent measurement are consistent with our hot carrier transport model. In particular, as is shown in Fig.~\ref{fig:fig2nn}(a) and Fig.~\ref{fig:fig3nn}(a), the zeros of the photocurrent for Landau levels $LL_{n\leq3}$  gradually shift away from the half-fillings for decreasing $n$. As seen in Fig.~4(a), this shift can be fit with an exponential suggesting the hot carrier distribution is present through several Landau levels. We also evaluate the sum of the photocurrent measured for +\textit{B} and -\textit{B} fields, and we observe oscillations with a relative high amplitude ($\sim$ 40~nA). This is shown in Fig.~4(b), and we note this observation cannot be explained in the photo-Nernst model where no oscillations are expected~\cite{Cao2016}. Finally, measurement at 4~T shows that oscillation amplitudes remain constant through several Landau levels--beyond a back gate voltage of 50~V (Fig. 2(a)), while data and theory for the Nernst effect suggest the photocurrent decays quickly away from the charge neutrality point.

\begin{figure*}
\begin{center}
\includegraphics[width = 0.8\linewidth]{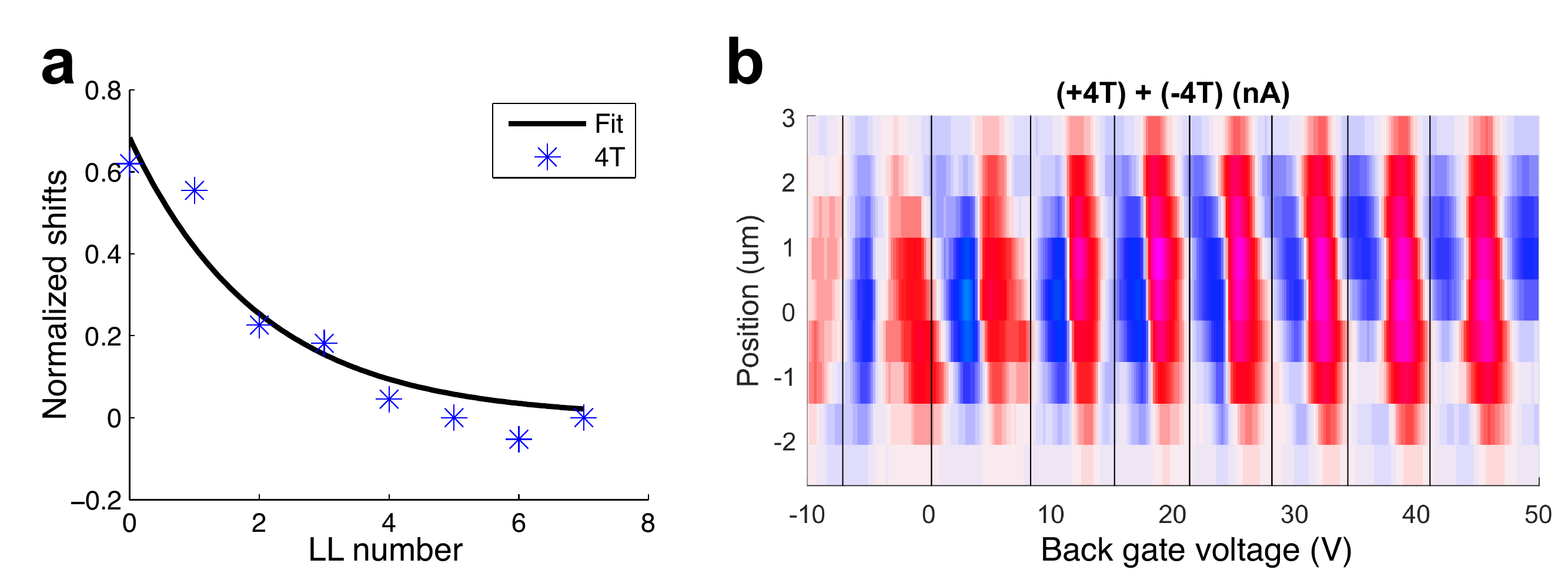}
\caption{\textbf{Non-chiral transport of the photo-carriers.} \textbf{(a).}  With a magnetic field of +4~T, the voltage difference between the zeros of the photocurrent and the half-fillings of Landau levels are shown, normalized by the width of individual Landau level. The difference approaches zero exponentially for increasing Landau level number, indicating the distribution of non-equilibrium carriers is present through several Landau levels.\textbf{(b).} The sum of the photocurrent measured at +4~T and -4~T as a function of back gate voltage shows oscillation.} 
\label{fig:fig4nn}
\end{center}
\end{figure*}

In summary, we present optical probing of a monolayer graphene in the quantum Hall regime through the generation of non-equilibrium carriers. Our photocurrent measurement provides deeper insights into the carrier transport behavior in the quantum Hall regime, in particular, their chirality above and below the Dirac point. Such optical probing permits the study of the quantum Hall states in an intermediate length-scale regime, which could be applied to other 2D material systems~\cite{Ju2017,Ma2017,Kastl2015,Arikawa2017a}. In addition, this work will contribute to developing novel applications of semiconductor materials, such as engineering the Peltier coefficient~\cite{Skinner2017}.

We acknowledge support by the NSF Physics Frontier center at the JQI, PFC@JQI. We also acknowledge fruitful discussions with Xiao Li, Wade DeGottardi and Fereshte Ghahari.

\appendix*
\section{Sample preparation and transport measurements}

\subsection{Sample fabrication}
Highly oriented pyrolytic graphite (HOPG) graphene and hexagonal boron nitride (hBN) are exfoliated using Scotch tape at ambient environment onto SiO2/Si substrate that is cleaned in Pirana solution followed by O2 plasma cleaning~\cite{Wang614}. The top hBN is picked up by a stacking of glass (0.25 mm thick, on top), PDMS (1 mm thick) and PPC (spin coated at 3000 RPM onto PDMS). PDMS was O2 plasma treated before spin coating to promote adhesion of PPC. The pick up is done at 40 C using a home made transfer stage. Same procedure is used to pick up subsequent graphene and bottom hBN. The stacking is then pressed against the target substrate (SiO2/Si) which is heated to 90 C. The glass/PDMS/PPC is then detached from the target substrate. The sample is then immersed in acetone overnight to remove residual PPC. EBL (30 keV) is used to define sample shape, using PMMA A4 as mask. The etching is done in in the plasma of O2 (10 sccm) and SF6 (40 sccm) at a chamber pressure of 200 mtorr for about 1 min (etching rate is ~ 20 nm/min). A second EBL is used to place contacts (Cr (5 nm)/Pd (10 nm)/Au (70 nm)) at the edges of the device. The device is then wire bonded using Au wires (25 um thick). The doped Si substrate is used as a back gate to control the carrier density in graphene. 

\begin{figure}
\begin{center}
\includegraphics[width = 1\linewidth]{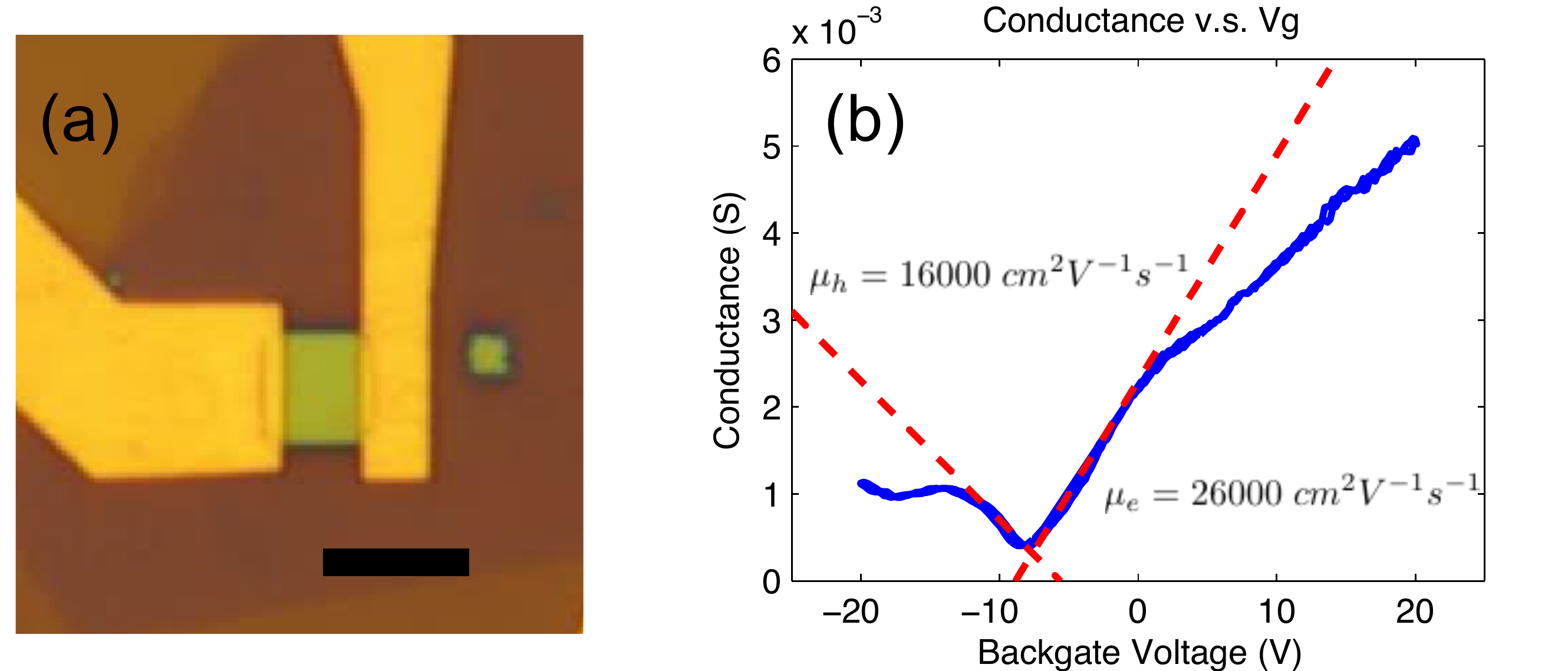}
\caption{(a) Microscope image of the sample. Scale bar is 5 $\mu$m. (b) Mobility measurement of the sample. The mobility for electrons is 26000 $cm^2V^{-1}s^{-1}$ and the mobility for holes is 16000 $cm^2V^{-1}s^{-1}$. The contact resistance is 121 $ \SI{}{\ohm}$.} 
\label{fig:S0nn}
\end{center}
\end{figure}

An optical microscope image of the fabricated sample is shown in Fig.~\ref{fig:S0nn}(a). The sample size is 2.49~$\mu$m by 3.87~$\mu$m. Measurements on other samples confirm the presented results. We estimate the contact resistance by calculating the sample resistance with the measured mobility. The difference between the measured resistance and the calculation is the contact resistance.

\subsection{Transport measurement under magnetic field}

\begin{figure}
\begin{center}
\includegraphics[width = 1\linewidth]{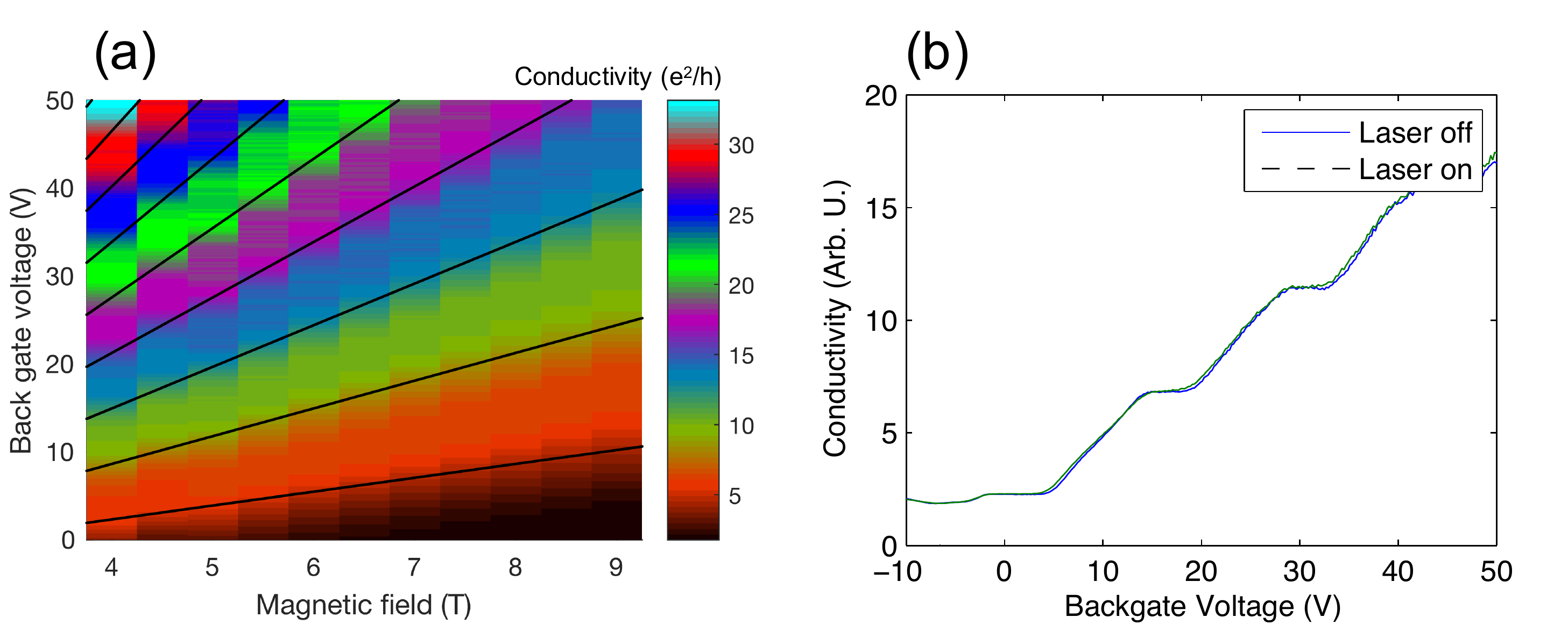}
\caption{(a) Fan diagram of the transport measurement under varying out-of-plane magnetic field and backgate voltage, $V_{\textrm{g}}$. The two-terminals conductivity shows Landau level quantization. The black lines are fitted Landau level half filling. (b) Comparison of two-terminal transport measurements at 9 T with laser on and off.
}
\label{fig:S1nn}
\end{center}
\end{figure}

Two-terminal transport measurements are conducted under various magnetic fields. Plateaus of conductance due to Landau level quantization can be readily seen, as shown in Fig. \ref{fig:S1nn} (a). In the transport measurement, an AC current of 1~$\mu$A at 13 Hz is injected through the drain and source contacts. A lock-in amplifier is locked at 13 Hz to measure the voltage drop between the drain and source. The conductance is obtained by taking the division between the 1~$\mu$A current and the voltage drop measured. During the measurement, a perpendicular static magnetic field is applied while the gate voltage is tuned to change the carrier density in the sample.

We compare the two-terminal transport measurements with laser on and off. We use an OPO laser with wavelength about 2 $\mu$m to excite the sample and a transport measurement is conducted. We turn off the laser and conduct another transport measurement. The comparison of the two transport measurements is shown in Fig.~\ref{fig:S1nn}(b). The two measurements overlap with each other very well showing no sign of heating.

\section{The photocurrent measurement}

We use near-infrared laser to excite electrons from below the Dirac point to above the Dirac point and measure the resulting photocurrent. We have repeated the photocurrent measurements on several samples. 

\subsection{The photocurrent measurement setup}

The sample, at 4.2~K, is mounted on  \textit{x, y, z} translation stages with relative position-readout. A free-space confocal microscope system is used to image the pump laser on the sample. 
The laser is a CW Ti:Sapphire laser of 10 $\mu$W with a fixed wavelength of 930 nm. This laser is chopped with a optical chopper at a frequency of 308 Hz. 

\begin{figure}[H]
\begin{center}
\includegraphics[width = 1\linewidth]{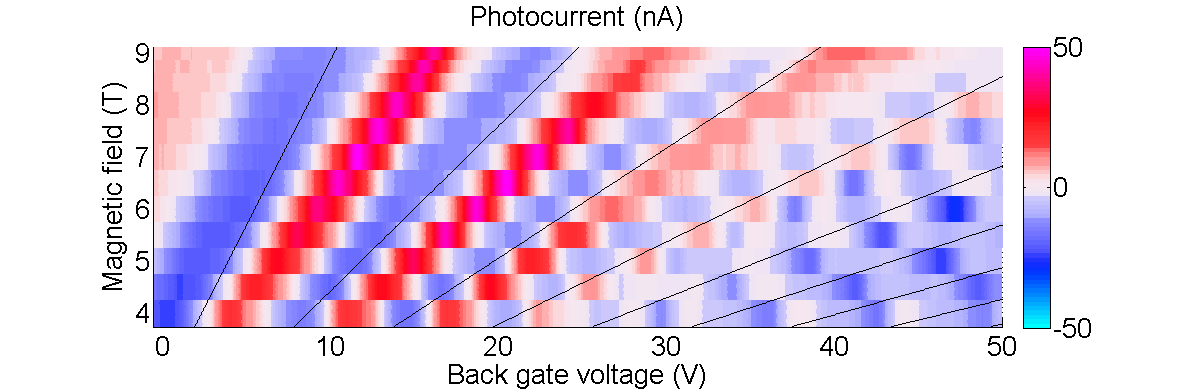}
\caption{Photocurrent measured in various magnetic fields. The polarity of the measured photocurrent shows a similar pattern as in the transport measurement. The black lines are the Landau level half-fillings as fitted for Fig.~\ref{fig:S1nn}.}
\label{fig:S2nn}
\end{center}
\end{figure}

The photocurrent is measured without any external bias. 
A lock-in amplifier, locked to the laser, is used to measure the laser generated photocurrent between the source and drain contacts.
During the measurement, both the perpendicular magnetic field and the backgate voltage are tuned. The resulting photocurrent as a function of magnetic field and backgate voltage shows a similar fan diagram as the transport measurement, as shown in Fig. \ref{fig:S2nn}.

\subsection{Photocurrent generation efficiency}

The 930 nm CW Ti:Sapphire laser corresponds to $\hbar\omega$=1.33~$eV$. The laser power applied on the surface of the sample is $P_{laser}$=10~$\mu W$. The number of photons per second in the pump laser is $n_{photon}=P_{laser}/\hbar\omega=4.8\times10^{13}~s^{-1}$. If we assume the absorption rate of a monolayer graphene is about 2\%~\cite{Wang2008} and to first order due only to particle-hole generation, the particle-hole generation rate, $n_{g}$ is then about $1\times10^{12}~s^{-1}=1~ps^{-1}$. 

The photocurrent amplitude we measure is in the regime of several tens of $nA$, which is equivalent to a rate of charged particles transport of $n_{e/h}=I_{photo}/e=6.2\times10^{10}~s^{-1}$. Comparing $n_{g}$ and $n_{e/h}$, we find that about 10\% of the excited particles and holes are measured in photocurrent. 

\subsection{Power dependence measurement}

The power dependence measurement shows that the laser power of 10 $\mu$W is in the linear response regime of the graphene photocurrent. Here, as an example, we show photocurrent oscillation as a function of back gate voltage for three laser powers: 2.5 $\mu$W, 5 $\mu$W, 10 $\mu$W, where it can be seen that doubling of the laser power results in doubling of the oscillation amplitude.

The linear response regime is also predicted by a simple, order of magnitude argument. The excited carrier lifetime has been measured~\cite{Plochocka2009}, leading to a recombination rate between $\rm 10^{13}-10^{14}~s^{-1}$. This exceeds by at least an order of magnitude the carrier generation rate ($\rm 10^{12}~s^{-1}$), and supports that the carrier generation at the laser power is in the linear response regime.

\begin{figure}
\begin{center}
\includegraphics[width = 0.8\linewidth]{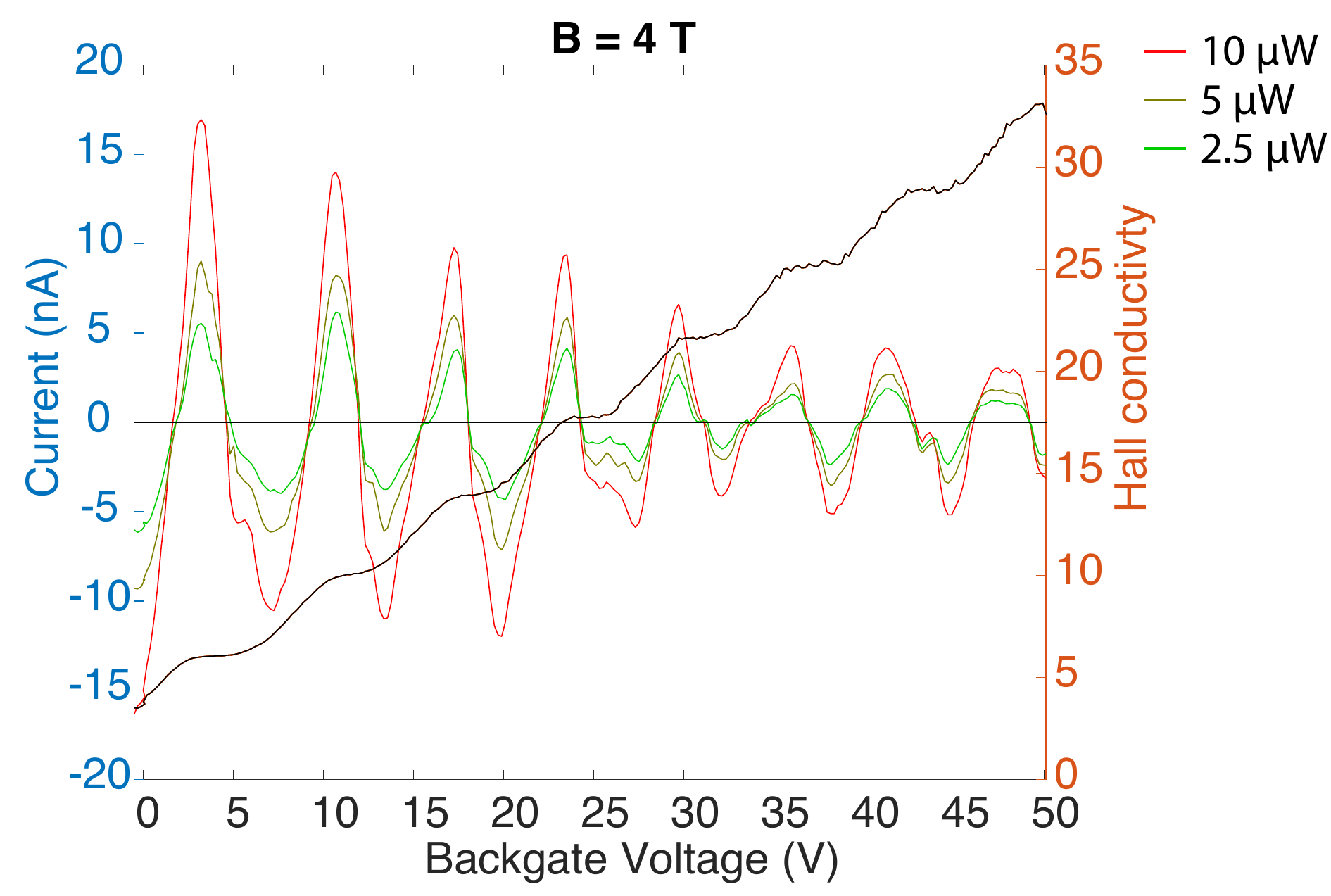}
\caption{Power dependence of the photocurrent and transport measurement at 4 T. The three lines shows photocurrent oscillation at different laser powers: 2.5 $\mu$W, 5 $\mu$W, 10~$\mu$W. Doubling of the laser power results in doubling of the photocurrent oscillation. Our measurement shows the power of 10~$\mu$W is in the linear response regime of graphene photocurrent. The black solid line is the conductance from the transport measurement.}
\label{fig:S3nn}
\end{center}
\end{figure}

\subsection{Laser spot size characterization}

The laser spot size is characterized by reflection from the gold contacts. The reflected light from the sample is collected through an optical fiber and measured. When the laser spot is on the gold contacts, the reflected power increases due to a higher reflection of gold as compared to the heterostructure. By moving the laser from graphene to a gold contact, the profile of the laser spot is mapped out. Our measurement shows the laser spot has a full-width half maximum of 1.8 $\mu$m.

\subsection{Photocurrent maps at magnetic fields of $\pm$4 T, $\pm$9 T}

We map out the photocurrent as a function of the position parallel to the contacts by moving the laser spot from edge to edge (along \textit{x}-direction in Fig.~1(c) in the main text), as shown in Fig. \ref{fig:S3nn}(a) for 4~T and Fig. \ref{fig:S3nn}(b) for -4~T. The laser is fixed at 10~$\mu$W and 930~nm. During the position scan, the laser spot is fixed in the center of the sample along y direction, perpendular to the two contacts. We observe photocurrent oscillations as a function of back gate voltage with polarity changes according to the Landau quantization. As explained in the main text, the photocurrent consists of diffusive current directly to the ohmic contacts and the edge state transport current. The polarity also switches in the center of the sample because now transport from the opposite edge begins to dominate. 

To obtain the edge state and bulk contribution, we sum over or subtract the $\pm 4~$T results~\cite{Cao2016}, as shown in Fig.~\ref{fig:S3nn}(c),(d). We observe that the edge transport component of the photocurrent is weak when the laser is focused in the center compared to the edge, and the two edges have opposite polarities, Fig.~\ref{fig:S3nn}(c). In contrast, the bulk transport part of the photocurrent is symmetric along the \textit{x}-direction, Fig.~\ref{fig:S3nn}(d).
Similar measurements are done for $\pm9~$T, as shown in Fig.~\ref{fig:S4nn}.

\begin{figure*}
\begin{center}
\includegraphics[width = 0.9\linewidth]{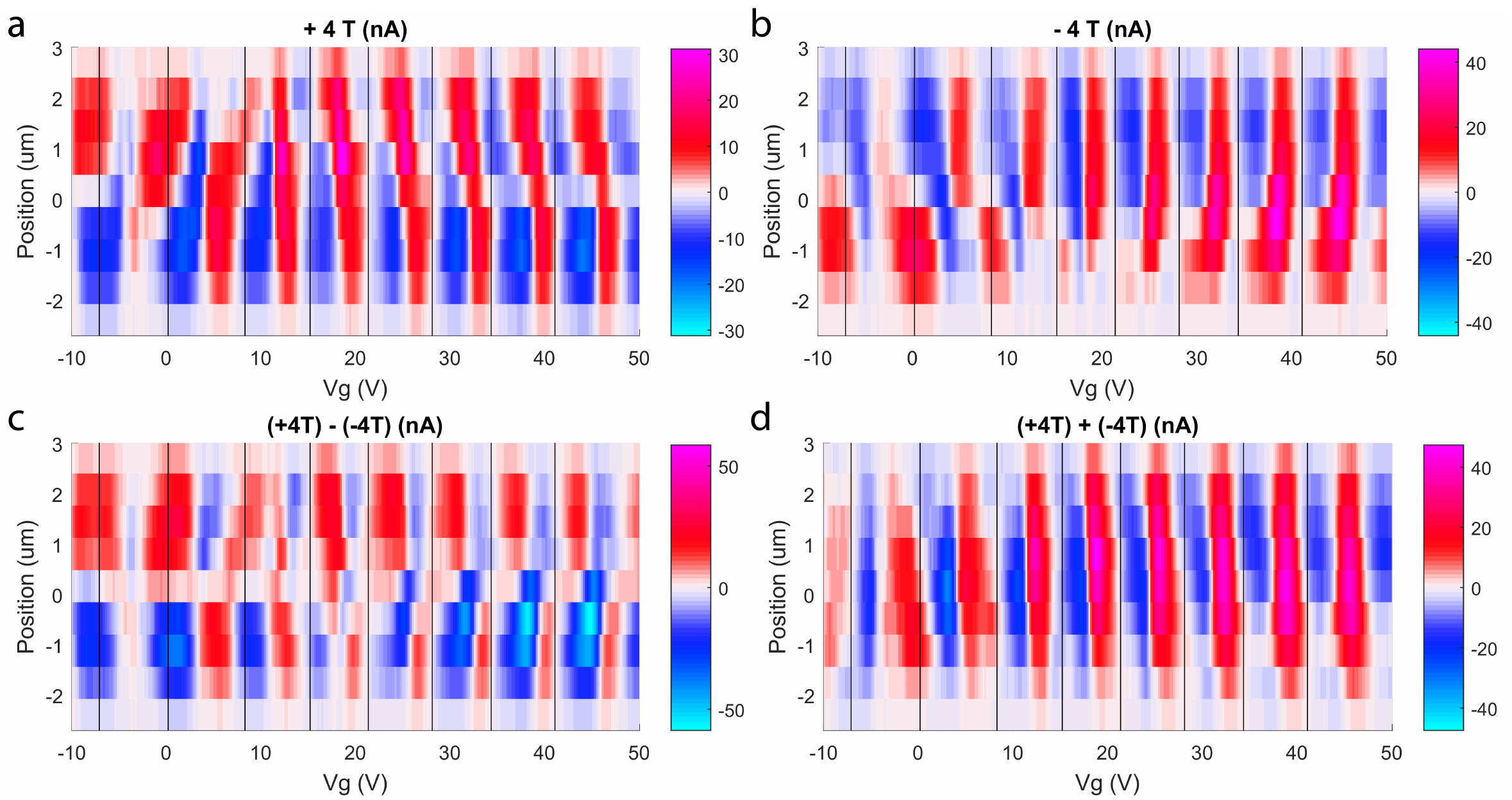}
\caption{Photocurrent maps as a function of back gate voltage for different laser positions. \textbf{(a),(b),} Photocurrent maps for +4~T and -4~T. \textbf{(c).} Difference of (a) and (b). \textbf{(d).} Sum of (a) and (b).}
\label{fig:S3nn}
\end{center}
\end{figure*}

\begin{figure*}
\begin{center}
\includegraphics[width = 0.9\linewidth]{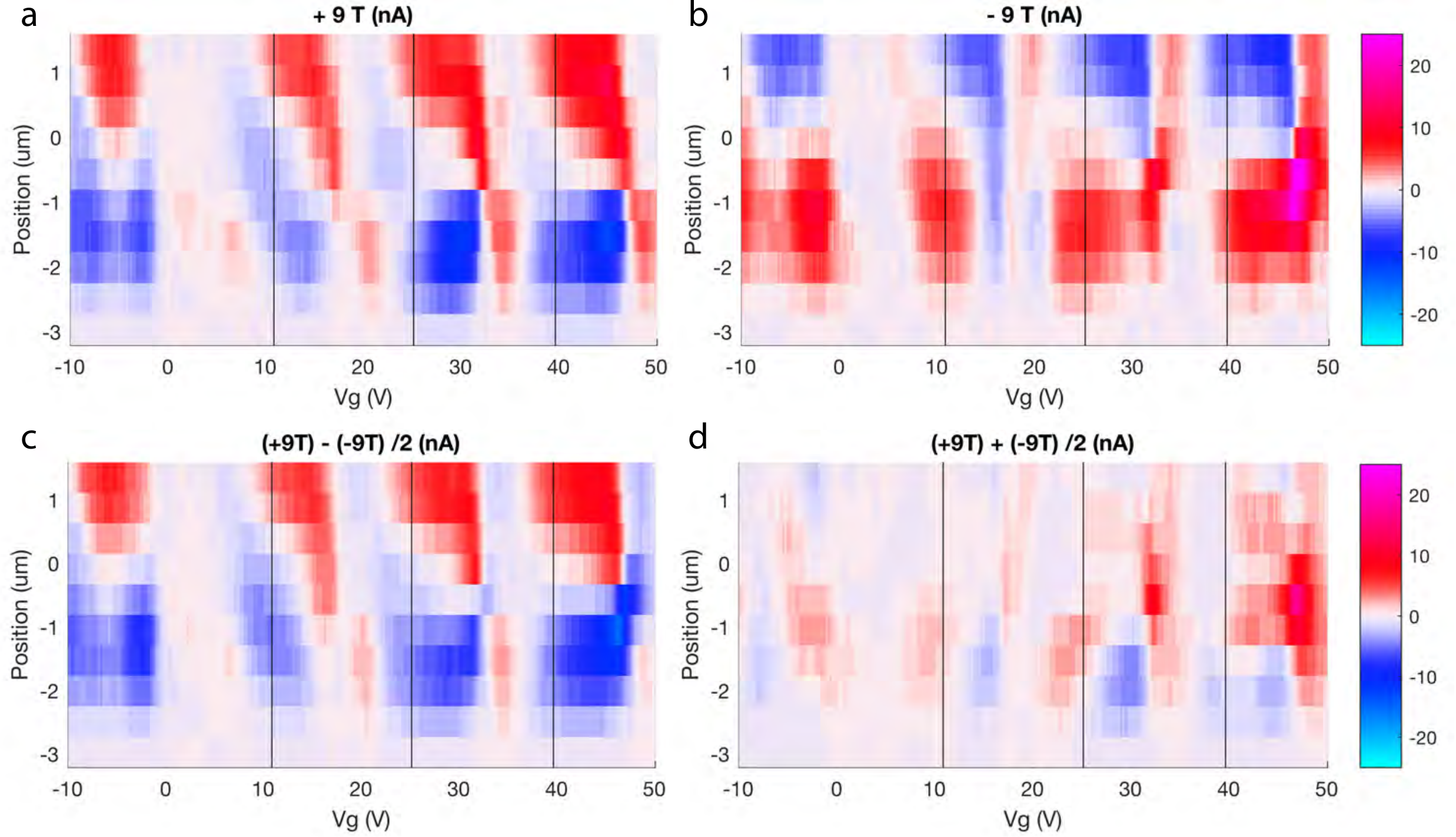}
\caption{Photocurrent maps as a function of back gate voltage for different laser positions. \textbf{(a),(b),} Photocurrent maps for +9~T and -9~T. \textbf{(c).} Difference of (a) and (b). \textbf{(d).} Sum of (a) and (b).}
\label{fig:S4nn}
\end{center}
\end{figure*}

\section{Photocurrent mechanism}

The explanation of the photocurrent generation can be separated into two parts: A. Imbalanced relaxation of hot carriers. B. Chirality of edge states. These two mechanisms in combination results in the photocurrent that we observe in the experiments.

\subsection{Imbalanced relaxation of hot carriers}

The relaxations of electrons and holes are imbalanced when the Fermi level is not in the half-filling of a Landau level. The relaxation rate of hot carriers is determined by the number of available states. When the Fermi level is above the half-filling, the number of available states for electrons is smaller than the one for holes resulting in that the holes relax faster than electrons. In this regime, the number of hot electrons is greater than the number of hot holes. These hot electrons/holes distribute in multiple Landau levels above/below the Fermi level. The hot carriers dissipate to the nearest edge and give a hot-electron-dominate photocurrent. A similar mechanism is described in~\cite{Nazin2010b}.

By sweeping the Fermi level through multiple Landau levels, the hot carrier type switches between electrons and holes multiple times, which manifests as the change of polarities of the measured photocurrent. However, the absolute type of the photocurrent's polarity is determined by the transport direction of the hot carriers.

\subsection{Chirality of edge states}

Aside from deriving the chirality of edge states from the Schr\"{o}dinger equation as introduced in the main text, we can also describe it with the Dirac equation~\cite{Queisser2013}. 

\begin{equation}
i\gamma^{\mu}(\partial_{\mu}+iA_{\mu})\Psi=0
\label{Eqn1}
\end{equation}

where $\Psi=(\psi_{1},\psi_{2})$ is the two-atom basis wavefunction
and we can assume $\psi_{1}$ is real and $\psi_{2}$ is imaginary.
$x^{\mu}=(v_{F}t,x,y)$ and $\gamma^{\mu}=(\sigma^{z},i\sigma^{y},-i\sigma^{x})$
are the Pauli matrices. $v_{F}$ is the Fermi velocity. We use the Landau gauge here $A_{\mu}=(0,0,xB\bf{y})$.
Because of the translation symmetry, we can separate variables as
$\Psi^{E,k}(t,x,y)=\exp(-iEt+iky)\Psi^{E,k}(x).$ Then we can have equations,

\begin{equation}
\begin{split}
iv_{F}[\partial_{\mu}+k+xB)]\psi_{2}^{E,k} & =E\psi_{1}^{E,k}\\
iv_{F}[\partial_{\mu}-k-xB]\psi_{1}^{E,k} & =E\psi_{2}^{E,k}
\label{Eqn2}
\end{split}
\end{equation}

The other set of solution $(\Psi^{E,k})^{*}$ can be obtained by substituting $E\rightarrow-E,\psi_{2}^{E,k}\rightarrow-\psi_{2}^{E,k}$. The way to interpret the two solutions $\Psi^{E,k}$ and $(\Psi^{E,k})^{*}$ is that $\Psi^{E,k}$ is the electron or hole with energy $E$ (which can be positive or negative). Correspondingly $(\Psi^{E,k})^{*}$ is the hole or electron with energy
$-E$~\cite{greiner2012quantum}.

The definition of the current~\cite{Queisser2013} is,
\begin{equation}
J=qv_F(\Psi^{E,k})^\dagger\gamma^0\gamma^\mu \Psi^{E,k}
\label{Eqn3}
\end{equation}
Here, $q$ is the carrier charge. Using eqn.~(\ref{Eqn2}), it is found that electron/hole can only flow in the \textit{y}-direction,

\begin{equation}
J^{y}=-\frac{2q v_{F}^{2}}{E}\int dx (\psi_{1}^{E,k})^2 k
\label{Eqn4}
\end{equation}

For electron and hole on the same side of the Dirac point \textit{i.e.} $E_{e} E_{h}>0$, the flow directions are the same for electron and hole, leading to opposite currents because of the opposite charge of electron and hole. For electron and hole separated by the Dirac point, \textit{i.e.} $E_{e} E_{h}<0$, the flow directions are opposite, giving additive currents. The edge state chirality is independent of the type of edge, zigzag or armchair~\cite{Abanin2007}.

The edge states transport plays a key role in the photocurrent generation. Hot electrons and holes diffuse to edge states and contribute to the photocurrent. The chirality of edge states, thus, determine the polarity of the photocurrent. In the high Landau level regime where all the hot electrons and hot holes are on the same side of the Dirac point, we see polarity change of photocurrent when the backgate voltage is at the half fillings of one Landau levels, as shown in the dashed box $E_F>0$ in Fig.~3(a) in the main text.

In the zeroth Landau level, the case is different. When electron and hole are separated by the Dirac point, the edge states transport to opposite directions which leads to the photocurrent of the same polarity, as is shown in the derivation above. Thus, at the zeroth Landau level, we see peaks of photocurrent which centers at the Dirac point, as shown in the dashed box $E_F=0$ in Fig.~3(a) in the main text. When the Fermi level moves away from the Dirac point to the high Landau level regime, the half-filling gradually shifts to coincide with the zeros of the photocurrent. This is due to the distribution of hot carriers through multiple Landau levels as explained above.

The mobility difference of electrons and holes results in the different photocurrent amplitudes of two polarities. We include this fact in our simulation.

\section{Model}

In our model, photocurrents are due to hot carriers which have not relaxed back to the Fermi energy. Relaxation rates $\Gamma_{\rm r}^{\rm (e)}$ ($\Gamma_{\rm r}^{\rm (h)}$) for electrons (holes) are determined by the density of states within a small strip above (below) the Fermi level, given through the gate voltage $V_{\textrm{g}}$. The Landau levels are centered around 
integer multiples of $V_0$, the voltage difference between a filled and an empty Landau level. Note that the relation between Fermi energy and gate voltage is $E_{\rm F} \sim \sqrt{V_{\textrm{g}}}$, leading to an equidistant spacing of Landau levels with respect to gate voltage. The number of states is assumed to be distributed according to a Lorentzian of width $\gamma$, so for the density of states we may write $\rho(V_{\textrm{g}})\sim \sum_n \frac{\gamma^2}{(V_{\textrm{g}}-nV_0)^2-\gamma^2}$. For our sample, we estimate $\gamma/V_0\approx 0.08$, and we consider a carrier as relaxed if the corresponding gate voltage is within $[V_{\textrm{g}}-4\gamma,V_{\textrm{g}}+4\gamma]$. Thus, for the relaxation rates, we write: 
\begin{align}
\label{Gamma}
\Gamma_{\rm r}^{\rm (e)}(V_{\textrm{g}}) &= \frac{1}{V_0t_{\rm r}} \int_{V_{\textrm{g}}}^{V_{\textrm{g}}+4\gamma} \rho({V_{\textrm{g}}}') {\rm d}{V_{\textrm{g}}}', \\ 
\Gamma_{\rm r}^{\rm (h)}(V_{\textrm{g}}) &= \frac{1}{V_0t_{\rm r}} \int_{V_{\textrm{g}}-4\gamma}^{V_{\textrm{g}}} \rho({V_{\textrm{g}}}') {\rm d}{V_{\textrm{g}}}',
\end{align}
In these expressions we have introduced a parameter $t_{\rm r}$ for the time scale of the relaxation process. The probability that an electron (e) or hole (h)  contributes to the photocurrent, $p_{\rm pc}^{\rm (e,h)}$, will depend on its relaxation rate $\Gamma_{\rm r}^{\rm (e,h)}$, and another time scale, $t_{\rm edge}$, the time which is needed for the carrier to reach the edge. We have: 
 \begin{align}
 \label{ppc}
  p_{\rm pc}^{\rm (e,h)}(V_{\textrm{g}}) = \frac{ 1/t_{\rm edge}}{  1/t_{\rm edge} +  \Gamma_{\rm r}^{\rm (e,h)}(V_{\textrm{g}}) }.
 \end{align}
From this expression, we determine the net photocurrent signal by considering the direction of edge transport. It is given by the chirality of the Landau level and the geometric edge which carries the charge. As argued in Eq. (1) of the main text, the chirality of a Landau level changes with the sign of its energy, and also opposite geometric edges have opposite direction of transport. 

\begin{figure}[t!]
\includegraphics[width=0.3\textwidth]{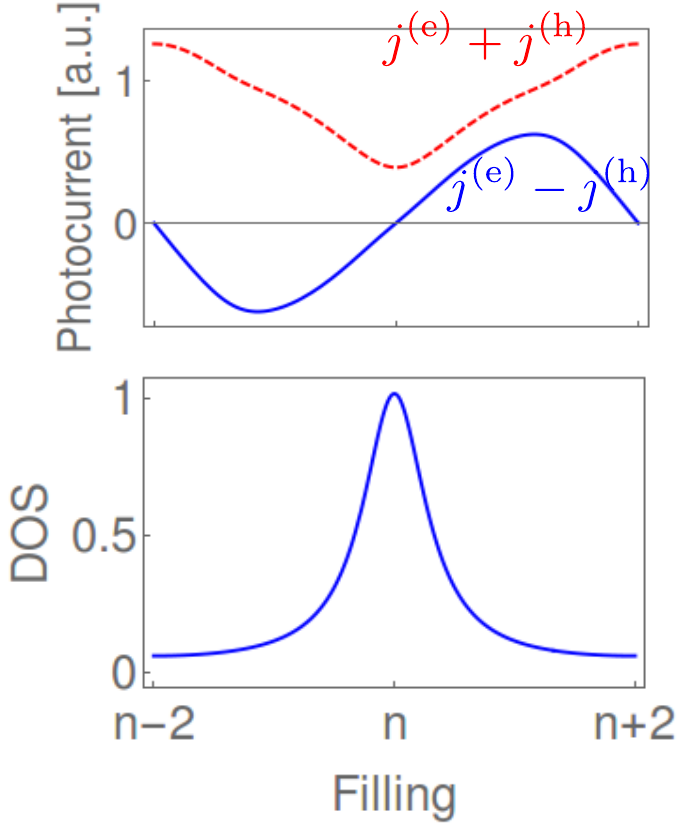}
\caption{\label{fig1}
Photocurrent (in arbitrary units) as a function of filling factor, obtained from the probabilities in Eq.~\ref{ppc} (using $t_{\rm e}=t_{\rm r}$). The red dashed curve considers counter-propagating electrons and holes, as expected near the Dirac point, and shows no polarity change.  The blue solid curve considers co-propagating electrons and holes, as expected near the Dirac point, and the current changes its polarity at half filling and integer filling. The density of states is shown in the lower panel.}
\end{figure}

For concreteness, let us now assume a positive Fermi energy. For the moment, let us also assume that only one geometric edge is involved in transport. With these assumptions, the transport direction for the electrons is fixed, but holes appear either with positive or negative energy, leading to transport either along with the electrons or in opposite direction.
 Thus, knowledge of the energy distribution of hot holes is needed in order to determine the photocurrent signal. Without that knowledge, we may still consider the two limiting cases:
One limiting case occurs when the Fermi level approaches the Dirac point, such that the vast majority of holes must be at negative energy. Then, electrons and holes are counter-propagating, and due to their opposite charges their photocurrent contributions add up: $j(V_{\textrm{g}}) \sim p_{\rm pc}^{\rm (e)}(V_{\textrm{g}}) + p_{\rm pc}^{\rm (h)}(V_{\textrm{g}})$. This results in a strong signal which does not display a polarity change along $V_{\textrm{g}}$, see red curve in upper panel of Fig.~\ref{fig1}.

Conversely, in the other limiting case, the Fermi level is far from the Dirac point. Then, almost all holes have relaxed to positive energies, since the relaxation through highly excited empty levels is much faster than relaxation processes near the Fermi level. In this case, electrons and holes move along the same direction, and their contributions to the photocurrents point in opposite directions, $j(V_{\textrm{g}}) \sim p_{\rm pc}^{\rm (e)}(V_{\textrm{g}}) - p_{\rm pc}^{\rm (h)}(V_{\textrm{g}})$. At the extrema of $\rho(V_{\textrm{g}})$,  the density of states is symmetric, and electron and hole contribution cancel each other. Thus, the polarity of the photocurrent changes between two Landau levels and at half-filled levels, see blue curve in upper panel of Fig.~\ref{fig1}.

\begin{figure}
\includegraphics[width=0.49\textwidth]{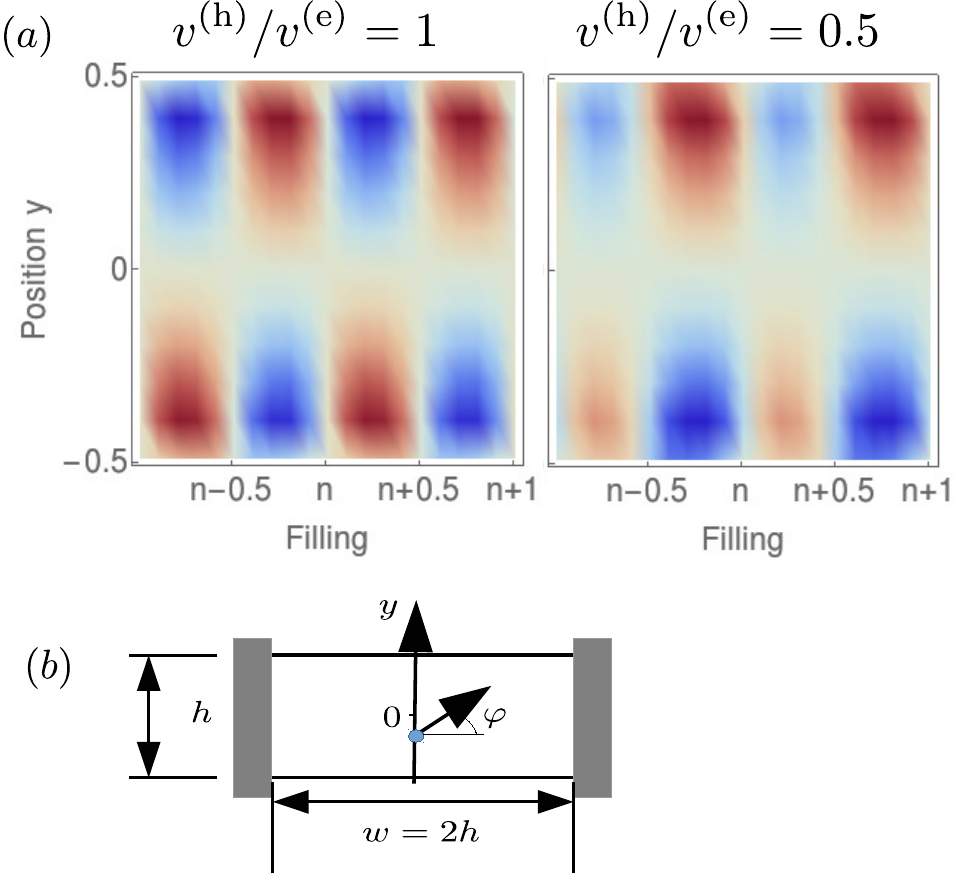}
\caption{\label{fig2}
(a) Photocurrent contour plots obtained from Eq.(\ref{pc}) as a function of filling and position of the laser. The left panel assumes equal mobilities for electrons and holes, for which positive (red) and negative (blue) photocurrent peaks are equally strong. For different mobilities shown in the right panel, one edge features stronger positive peaks, while the other edge is dominated by negative peaks. (b) Geometry of the sample: Carries are produced along the $y$-axis, and drift in an arbitrary direction $\varphi$.}
\end{figure}

In the following, we will take into account the existence of two geometric edges. This becomes important when carriers are not excited in close vicinity to one edge. The photo-excited carriers are expected to drift in an arbitrary direction, and may reach any edge of the sample, or even one of the leads without going through an edge channel. Position $y$ of a carrier, and its drift direction $\varphi$ (cf. Fig.~\ref{fig2}(b)) define the distance $d(y,\varphi)$ to the edge. The drift time is given by $t_{\rm edge}^{\rm (e,h)}=d(y,\varphi)/v^{\rm (e,h)}$, where $v^{\rm (e,h)}$ is the drift velocity. Here, we assume different velocities for electrons and holes due to different mobilities. Substituting $t_{\rm edge}$ in Eq.(\ref{ppc}) by these expressions, we obtain a photocurrent probability $p_{\rm pc}^{\rm (e,h)}(V_{\textrm{g}},y,\varphi)$ which depends not only on the Fermi level $V_{\textrm{g}}$, but also on the position $y$ and the drift direction $\varphi$. To keep track of the polarity of the corresponding current, we also introduce a binary variable $\sigma(y,\varphi)=\pm 1$, where the sign is defined by the geometric edge which is reached by a drift from $y$ along $\varphi$.

In the two limiting cases with all carriers of one species having the same chirality, the photocurrent contribution of each carrier type is then given by 
 \begin{align}
 \label{pc}
  j(y,V_{\textrm{g}})^{\rm (e,h)} \sim \int_{0}^{2\pi} \sigma(y,\varphi)  p_{\rm pc}^{\rm (e,h)}(V_{\textrm{g}},y,\varphi) {\rm d}\varphi.
 \end{align}
For example, sufficiently far away from the Dirac point, the total photocurrent signal will then be given by $j(y,V_{\textrm{g}})=j(y,V_{\textrm{g}})^{\rm (e)}-j(y,V_{\textrm{g}})^{\rm (h)}$, as plotted in Fig.~\ref{fig2}(a) for equal and unequal mobilities of electrons and holes ($v^{\rm (e)}/v^{\rm (h)}=1$ and $v^{\rm (e)}/v^{\rm (h)})=2$). Drift velocities have been related to the relaxation time scale $t_{\rm r}$ introduced in Eq. (\ref{Gamma}) by $v^{\rm (e)}=0.2 \ \mu{\rm m}/t_{\rm r}$.

Finally, in order to model the full experimental data, we must take into account that holes (electrons) of any chirality co-exist when the Fermi energy is above (below) the Dirac point. Therefore, we introduce a Boltzmann weight, and set the gate voltage at the Dirac point to zero, $V_{\textrm{g}}=0$. Then, the portion of electrons (holes) below (above) the Dirac point is assumed to be given by $e^{-\beta |V_{\textrm{g}}|}$ for $V_{\textrm{g}}<0$ ($V_{\textrm{g}}>0$),  with $\beta$ being a temperature-like parameter. For the total photocurrent signal we then write:
\begin{widetext}
\begin{align}
\label{pc2}
   j(y,V_{\textrm{g}}) &\sim j(y,V_{\textrm{g}},+1)^{\rm (e)} - e^{-\beta |V_{\textrm{g}}|} j(y,V_{\textrm{g}},-1)^{\rm (h)} - (1-e^{-\beta |V_{\textrm{g}}|} ) j(y,V_{\textrm{g}},+1)^{\rm (h)}  \ \ \ \ \ {\rm if} \ V_{\textrm{g}}\geq0, \nonumber \\
   j(y,V_{\textrm{g}}) &\sim (1-e^{-\beta |V_{\textrm{g}}|} ) j(y,V_{\textrm{g}},+1)^{\rm (e)} + e^{-\beta |V_{\textrm{g}}|} j(y,V_{\textrm{g}},-1)^{\rm (e)} - j(y,V_{\textrm{g}},-1)^{\rm (h)} \ \ \ \ \ {\rm if} \ V_{\textrm{g}}<0.  \\
\end{align}
\end{widetext}
Our modeling in the main text, shown in Fig.~3(d), was obtained from Eq. (\ref{pc2}) using the parameters $v^{\rm (e)}=0.2 \mu{\rm m}/t_{\rm r}, v^{\rm (h)}=0.1 \mu{\rm m}/t_{\rm r}, \beta=1/V_0$. The laser beam size of full-width-half-maxium 1.8~$\mu m$ is taken into the consideration by convolution. The little dip in the photocurrent intensity at the zeroth Landau level is due to the fast electron-hole recombination rate.

\end{document}